\documentclass[journal]{IEEEtran}

\usepackage{booktabs}
\usepackage{amssymb}
\usepackage{cite}

\usepackage{color,soul} 

\ifCLASSINFOpdf
  \usepackage[pdftex]{graphicx}
\else
\fi
\usepackage{amsmath}
\usepackage{url}

\begin{document}
%
\title{An Instrumental Intelligibility Metric Based \\ on Information Theory}
%
%
%


\author{Steven~Van~Kuyk,~\IEEEmembership{Student~Member,~IEEE,}
        W.~Bastiaan~Kleijn,~\IEEEmembership{Fellow,~IEEE,}
        and~Richard~C.~Hendriks,~\IEEEmembership{Member,~IEEE}


\thanks{Copyright (c) 2017 IEEE. Personal use of this material is permitted. However, permission to use this material for any other purposes must be obtained from the IEEE by sending a request to pubs-permissions@ieee.org.}
\thanks{S. Van Kuyk is with Victoria University of Wellington, New Zealand (email:steven.van.kuyk@ecs.vuw.ac.nz).}
\thanks{W. B. Kleijn is with Victoria University of Wellington, New Zealand and also with Delft University of Technology, The Netherlands (email: bastiaan.kleijn@ecs.vuw.ac.nz).}
\thanks{R. C. Hendriks is with Delft University of Technology, The Netherlands (email: r.c.hendriks@tudelft.nl).}
}

%
%

\markboth{Accepted for publication in IEEE SIGNAL PROCESSING LETTERS, DOI: 10.1109/LSP.2017.2774250}%
{}
%



\maketitle

\begin{abstract}
We propose a monaural intrusive instrumental intelligibility metric called SIIB (speech intelligibility in bits). SIIB is an estimate of the amount of information shared between a talker and a listener in bits per second. Unlike existing information theoretic intelligibility metrics, SIIB accounts for talker variability and statistical dependencies between time-frequency units. Our evaluation shows that relative to state-of-the-art intelligibility metrics, SIIB is highly correlated with the intelligibility of speech that has been degraded by noise and processed by speech enhancement algorithms.
\end{abstract}


%
\IEEEpeerreviewmaketitle

\section{Introduction}
%
%
%
%
\IEEEPARstart{I}{ntelligibility} is defined as the proportion of words correctly identified by a listener and is a natural measure for quantifying the effectiveness of speech-based communication systems \cite{allen2005articulation}. Although listening tests can provide valid data, such tests are time-consuming to conduct. For this reason, instrumental intelligibility metrics that are correlated with intelligibility and quick to compute are often preferred.

We can distinguish two types of instrumental intelligibility metrics: intrusive, and non-intrusive. Intrusive intelligibility metrics require knowledge of the clean speech and either the communication channel or degraded speech, whereas non-intrusive intelligibility metrics require only the degraded speech. In this paper we develop a new intrusive intelligibility metric based on information theory \cite{shannon1948mathematical}.

{Existing intrusive intelligibility metrics} include the speech intelligibility index (SII) \cite{SII12}, the speech transmission index (STI) \cite{houtgast1971evaluation}, the coherence SII (CSII) \cite{kates2005coherence}, the extended SII (ESII) \cite{rhebergen2005speech}, 
the normalized covariance measure (NCM) \cite{Koch1992Auditory, goldsworthy2004analysis}, the hearing-aid speech perception index (HASPI) \cite{kates2014hearing}, {the short-time objective intelligibility measure (STOI)}\cite{taal2011algorithm}, {the extended STOI (ESTOI)}\cite{jensen2016algorithm}, {the speech-based envelope power spectrum model (sEPSM)} \cite{jorgensen2011predicting,jorgensen2013multi,relano2016predicting}, {and the glimpse proportion metric (GP)} \cite{cooke2006glimpsing,barker2007modelling,tang2016glimpse}. As a group, the above algorithms have been successful at predicting speech intelligibility in a wide-range of conditions including additive noise, filtering, reverberation, and non-linear enhancement. However, each intelligibility metric tends to perform well for only a narrow subset of conditions. This is because the above algorithms were heuristically motivated and were often designed with a specific type of distortion or data set in mind. 

Information theory provides a mathematical framework for modelling communication systems. Information theoretical concepts have previously been used in the analysis of linguistics \cite{shannon1951prediction,pellegrino2011across}, speech production \cite{kleijn2015simple}, and human hearing \cite{smith2006efficient}. Additionally, state-of-the-art speech enhancement algorithms \cite{kleijn2015simple,khademi2017intelligibility} and intelligibility metrics \cite{taghia2014objective, jensen2014speech, van2016intelligibility} that are based on information theory have been developed.

{Existing information theoretic intelligibility metrics, such as the mutual information k-nearest neighbour metric (MIKNN)} \cite{taghia2014objective}, assume that speech can be described by a memoryless stochastic process and that the energy of a speech signal at one time-frequency location is statistically independent to the energy at all other time-frequency locations. In reality neither of these assumptions are valid, which leads to an over-estimate of the information shared between a talker and a listener.

In this paper we propose a conceptually simple intelligibility metric called SIIB. SIIB is a function of a clean acoustic signal produced by a talker and a degraded signal that is received by a listener. As described in Section II and Section III, the acoustic signals are converted to a representation of speech based on a crude model of the human auditory system. A non-parametric estimate of the mutual information rate of the signals is then computed. Unlike existing metrics, SIIB partially accounts for time-frequency dependencies in the speech signals using the Karhunen-Lo\`eve transform (KLT) \cite{karhunen1947lineare} and incorporates the theory developed in \cite{kleijn2015simple} to account for the effect that talker-variability has on the information rate. In Section IV and Section V, SIIB is evaluated by comparing its performance to STOI \cite{taal2011algorithm}, ESTOI \cite{jensen2016algorithm}, and MIKNN \cite{taghia2014objective} for speech degraded by noise and processed by enhancement algorithms.

\section{Model of Speech Communication}
In this section we present a theoretical model of speech communication similar to that described in \cite{kleijn2015simple, van2016intelligibility}, and \cite{van2017onthe}. The model considers the transmission of a message from a talker to a listener. Stochastic processes are denoted by $\{\cdot \}$, random variables are denoted by bold font, and their realisations are denoted by regular font. 

\subsection{The Communication Channel}
A message $\{\mathbf{M}_t\}$, speech signal $\{\mathbf{X}_t\}$, and degraded speech signal $\{\mathbf{Y}_t\}$ are represented by ergodic stationary discrete-time vector-valued random processes where $t\in \mathbb{Z}$ is the time index. The message can be thought of as a sequence of latent variables that represent, for example, a sequence of sentences, phonemes, or neural states. The talker encodes the message into a speech signal according to a conditional probability distribution $p_{\{\mathbf{X}_t\}|\{\mathbf{M}_t\}}(\{X_t\}|\{M_t\})$. In this way, the variability of different talkers encoding the same message into a speech signal is incorporated into the model. 

The speech signal is transmitted to a listener through a communication channel that may distort the signal. Examples of distortion include noise, reverberation, speech coding algorithms, and speech enhancement algorithms. Overall, the communication process is described by a Markov chain:
\begin{equation}
	\{\mathbf{M}_t\} \to \{\mathbf{X}_t\}  \to \{\mathbf{Y}_t\}. 
\end{equation}
We call $\{\mathbf{M}_t\} \to \{\mathbf{X}_t\}$ the speech production channel and $\{\mathbf{X}_t\}  \to \{\mathbf{Y}_t\}$ the environmental channel.

The representation of speech used in this paper is based on a crude model 
of the human auditory system and was motivated using information theoretic arguments in \cite{smith2006efficient} and \cite{van2017onthe}. Let $\{\mathbf{x}_i\}$ be a real-valued random process that represents the samples of an acoustic speech signal where $i$ is the sample index and let $\{\hat{\mathbf{x}}_t\}$ be the short-time Fourier transform (STFT) of $\{\mathbf{x}_i\}$ where $t$ is the frame index. We define $\mathbf{X}_t$ as an $\mathbb{R}^J$-valued random variable that represents auditory log-spectra given by \vspace*{-0.6mm}
\begin{equation}
	\mathbf{X}_t =  \ln G |(\hat{\mathbf{x}}_t)|^2,
	\label{eq:representation}
\end{equation}
where $G\in \mathbb{R}^{J\times N}$ is a matrix that represents an auditory filterbank, and the logarithm and squared magnitude operators are applied elementwise. To account for temporal masking in the auditory system, 
the masking function described in \cite{rhebergen2006extended} is applied to $\mathbf{X}_t$. The degraded speech $\mathbf{Y}_t$ is defined similarly. 

\vspace*{-0.5mm}
\subsection{Information Rate of the Communication Channel}
The proposed intelligibility metric is based on the hypothesis that intelligibility is a function of the mutual information rate between the message and the degraded speech. Let $\mathbf{M}^K = [(\mathbf{M}_1)^T, (\mathbf{M}_2)^T, \cdots, (\mathbf{M}_{K})^T ]^T$, where $T$ denotes the transpose, be a vector obtained by stacking $K$ consecutive message vectors and similarly for $\mathbf{Y}^K$. The mutual information rate is defined by
\begin{equation}
	I(\{\mathbf{M}_t\}; \{\mathbf{Y}_t\})=\lim_{K\to \infty} \dfrac{1}{K} I(\mathbf{M}^K;\mathbf{Y}^K), 
	\label{eq:information_rate}
\end{equation}
where $I(\mathbf{M}^K; \mathbf{Y}^K)$ is the mutual information between $\mathbf{M}^K$ and $\mathbf{Y}^K$ given by
\begin{equation}
\begin{aligned}
	I(\mathbf{M}^K; &\mathbf{Y}^K) = \\
& \int\limits_{\mathbf{M}^K, \mathbf{Y}^K}\hspace{-4pt} p(M^K, \hspace{-1pt}Y^K) \log_2 \dfrac{ p(M^K, Y^K)}{ p(M^K)p(Y^K)} dM^K dY^K.
\end{aligned}
\vspace*{-0.2mm}
\end{equation}

To estimate \eqref{eq:information_rate}, realisations of $\mathbf{M}_t$ and $\mathbf{Y}_t$ are needed. Estimating a realisation of $\mathbf{M}_t$ requires a chorus of speech signals (see \cite{van2017onthe}). In typical applications of intelligibility prediction, such a chorus is not available, so instead we use an upper bound on \eqref{eq:information_rate}. By applying the data processing inequality twice we have \cite{cover2012elements} 
\begin{equation}
	I(\{\mathbf{M}_t\};\{\mathbf{Y}_t\}) \leq \min \big(I(\{\mathbf{M}_t\};\{\mathbf{X}_t\}),I(\{\mathbf{X}_t\};\{\mathbf{Y}_t\})\big).
	\label{eq:bound}
\end{equation}
In the case of a distortionless environmental channel, $I(\{\mathbf{X}_t\};\{\mathbf{Y}_t\})$ is unbounded from above and $I(\{\mathbf{M}_t\};\{\mathbf{Y}_t\})$ saturates at the information rate of the speech production channel \cite{kleijn2015simple}. 
%
This maximum information rate is determined by the variability in pronunciation between different talkers.
%
%
The following subsections describe how $I(\{\mathbf{M}_t\};\{\mathbf{X}_t\})$ and $I(\{\mathbf{X}_t\};\{\mathbf{Y}_t\})$ can be calculated.

\subsection{Information Rate of the Environmental Channel}
The mutual information rate of the environmental channel is given by
\begin{equation}
	I(\{\mathbf{X}_t\};\{\mathbf{Y}_t\}) = \lim_{K\to\infty}\dfrac{1}{K}I(\mathbf{X}^K;\mathbf{Y}^K).
	\label{eq:I(XY)}
\end{equation}
Estimating the mutual information between vectors of high dimensionality is a challenging task \cite{doquire2012comparison} particularly when the vector elements have strong statistical dependencies \cite{gao2015efficient}. For this reason we introduce an invertible transform $f(\cdot)$ that aims to remove the dependencies between the vector elements. 

Let $\tilde{\mathbf{X}}^K = f(\mathbf{X}^K)$ and $\tilde{\mathbf{Y}}^K = f(\mathbf{Y}^K)$. {In the following we assume that the elements of $\tilde{\mathbf{X}}^K$ can be approximated as statistically independent}, and likewise for $\tilde{\mathbf{Y}}^K$. Then \eqref{eq:I(XY)} can be decomposed into a summation:
\begin{equation}
\begin{aligned}
	I(\{\mathbf{X}_t\};\{\mathbf{Y}_t\}) &= \lim_{K\to\infty}\dfrac{1}{K}I(\mathbf{X}^K;\mathbf{Y}^K)\\
	&= \lim_{K\to\infty}\dfrac{1}{K}I(\tilde{\mathbf{X}}^K;\tilde{\mathbf{Y}}^K)\\
	&= \lim_{K\to\infty}\dfrac{1}{K}\sum_{j=1}^{KJ} I(\tilde{\mathbf{X}}^K_j;\tilde{\mathbf{Y}}^K_j),
	\label{eq:Isum}
\end{aligned}
\end{equation}
where $j$ denotes the element index in the vector.

Finding an invertible $f(\cdot)$ that simultaneously removes the dependencies in both $\mathbf{X}^K$ and $\mathbf{Y}^K$ is difficult. Early speech recognition systems used the discrete cosine transform (DCT), which results in Mel-frequency cepstral coefficients \cite{davis1980comparison}. It can be shown that the DCT approximates the Karhunen-Lo\`eve transform (KLT) for stationary signals \cite{rao2014discrete}. The KLT is the transformation that we use here and it is given by:
\begin{equation}
	\tilde{\mathbf{X}}^K = U(\mathbf{X}^K-\mathrm{E}[\mathbf{X}^K]),
	\label{eq:KLT1}
\end{equation}
and
\begin{equation}
	\tilde{\mathbf{Y}}^K = U(\mathbf{Y}^K-\mathrm{E}[\mathbf{Y}^K]),
	\label{eq:KLT2}
\end{equation}
where $U$ is a matrix with rows equal to the unit-magnitude eigenvectors of the covariance matrix of $\mathbf{X}^K$ and $\mathrm{E}[\cdot]$ is the expected value operator. The KLT ensures that the elements of $\tilde{\mathbf{X}}^K$ are statistically uncorrelated, and if $\mathbf{X}^K$ is Gaussian, which is a reasonable approximation, then the elements are also statistically independent. 

The KLT does not guarantee the same properties for $\tilde{\mathbf{Y}}^K$ unless $\mathbf{Y}^K$ is also Gaussian and has a covariance matrix equal to a scalar multiple of the covariance matrix of $\mathbf{X}^K$. In practice the environmental channel can result in non-Gaussian $\mathbf{Y}^K$ or can introduce statistical dependencies in $\mathbf{Y}^K$ that are not present in $\mathbf{X}^K$. An example of the latter is a reverberant channel. In this case, the statistical dependencies in the source are accounted for by the KLT, but the statistical dependencies in the received signal are not accounted for. The consequence is that \eqref{eq:Isum} underestimates the mutual information rate. Although the KLT does not meet all of the requirements for $f(\cdot)$ we found that it improves performance.

\subsection{Information Rate of the Speech Production Channel}
{Approximating $\{\mathbf{M}_t\}$ and $\{\mathbf{X}_t\}$ as Gaussian}, the information rate of the speech production channel is
\begin{equation}
\begin{aligned}
	I(\{\mathbf{M}_t\};\{\mathbf{X}_t\}) &= \lim_{K\to\infty} \dfrac{1}{K} \sum_{j=1}^{KJ} I(\tilde{\mathbf{M}}^K_j;\tilde{\mathbf{X}}^K_j)\\
	&= \lim_{K\to\infty} -\dfrac{1}{K} \sum_{j=1}^{KJ} \dfrac{1}{2}\log_2 (1-r^2_j),
	\label{eq:rate_production}
\end{aligned}
\end{equation}
where $\tilde{\mathbf{M}}^K$ is defined similarly to $\tilde{\mathbf{X}}^K$ and $r_j$ is called the production noise correlation coefficient. The production noise correlation coefficient describes the efficiency of encoding a message into a speech signal according to $p_{\{\mathbf{X}_t\}|\{\mathbf{M}_t\}}(\{X_t\}|\{M_t\})$. Based on the measurements in \cite{van2016intelligibility} and \cite{van2017onthe}, this paper uses $r_j = 0.75$ for all $j$.

\section{Proposed Intelligibility Metric}
The proposed intelligibility metric combines \eqref{eq:Isum}, \eqref{eq:rate_production}, and \eqref{eq:bound} to give an estimate of the amount of information shared between $\{\mathbf{M}_t\}$ and $\{\mathbf{Y}_t\}$ in bits per second. It is given by
\begin{equation}
	\mathrm{SIIB} = \dfrac{F}{K} \sum_{j=1}^{KJ} \min\bigg(\hspace{-1mm}-\frac{1}{2}\log_2(1-r_j^2),\; {I}(\tilde{\mathbf{X}}^K_j;\tilde{\mathbf{Y}}^K_j) \bigg),
\end{equation}
where $F$ is the frame rate in Hz.

We now describe our implementation. An estimate of $I(\tilde{\mathbf{X}}^K_j;\tilde{\mathbf{Y}}^K_j)$ is computed by applying a k-nearest neighbour mutual information estimator \cite{kraskov2004estimating} to observed sample sequences $\tilde{X}^K_{j,t}$ and $\tilde{Y}^K_{j,t}$. To obtain $\tilde{X}^K_{j,t}$ and $\tilde{Y}^K_{j,t}$, a clean acoustic speech signal and a degraded signal are resampled to a sampling rate of 16 kHz. An energy-based voice activity detector with a 40 dB threshold is applied to remove silent segments. Subsequently, the signals are transformed to the STFT domain using a 400-point Hann window with 50\% overlap. This gives a frame rate of $F=80$ Hz, which is sufficient for capturing the spectral modulations required for high intelligibility \cite{elliott2009modulation}.

A gammatone filterbank \cite{slaney1993efficient} that includes $J=28$ filters linearly spaced on the ERB-rate scale \cite{glasberg1990derivation} between 100 Hz and 6500 Hz is used to obtain $X_t$ and $Y_t$ according to \eqref{eq:representation}. A sequence of stacked vectors for the clean speech is then formed by stacking $K=15$ consecutive vectors:
\begin{equation}
	X^K_t = [(X_{t-K+1})^T, (X_{t-K+2})^T, \cdots, (X_{t})^T ]^T
\end{equation}
and similarly for $Y^K_t$. Setting $K=15$ means that dependencies spanning 187.5 ms are considered. For comparison, the mean duration of a phoneme is 80 ms \cite{crystal1988segmental}. The sample covariance matrix of $X^K_t$ is computed and the KLT in \eqref{eq:KLT1} and \eqref{eq:KLT2} is applied to obtain $\tilde{X}_t^K$ and $\tilde{Y}_t^K$.

\section{Evaluation Procedures}
This section describes the procedures used to evaluate SIIB. The evaluation considered four intelligibility data sets and used \newpage \noindent two performance measures to quantify the strength of the relationship between SIIB and intelligibility. 

\subsection{Intelligibility Data Sets}

\subsubsection{JensenSCNR}
The first data set consists of speech subjected to single channel noise reduction. In \cite{jensen2012spectral} phrases from the Dutch version of the Hagerman test \cite{koopman2007development,hagerman1982sentences} were degraded by speech-shaped noise (SSN) at SNRs of $-8, -6, -4, -2,$ and $0$ dB and processed {by three} noise reduction algorithms. The three algorithms compute a minimum mean-squared error estimate of the clean speech by multiplying the short-time spectral magnitude of the degraded speech with a gain function. In total there are 5 SNRs $\times$ (3 algorithms + 1 unprocessed) = 20 conditions. The stimuli were presented to {13} normal-hearing subjects for identification.
\subsubsection{KleijnPRE}
The second data set consists of {speech subjected} to pre-processing enhancement {and degraded} by noise. In \cite{kleijn2015simple} {phrases} from the Dutch version of the Hagerman test were subjected {to three} pre-processing enhancement algorithms and then degraded either by SSN at SNRs of $-15, -12, -9,$ and $-6$ dB, or car noise at SNRs of $-23, -20, -17,$ and $-14$ dB. The three enhancement algorithms optimally redistribute the energy of the clean speech according to a distortion criterion. In total there are 2 noise types $\times$ 4 SNRs $\times$ (3 algorithms + 1 unprocessed) = 32 conditions. The stimuli were presented to nine normal-hearing listeners for identification.
\subsubsection{CookePRE}
The third data set also consists of speech subjected to pre-processing enhancement. In \cite{cooke2013intelligibility} {Harvard sentences} \cite{rothauser1969ieee} were processed {by 19} pre-processing enhancement algorithms and degraded either by SSN at SNRs of $1, -4,$ and $-9$ dB, or by speech from a competing talker at SNRs of $-7, -14,$ and $-21$ dB. The stimuli were presented to 175 normal-hearing listeners for identification. For this paper, a subset of the data in \cite{cooke2013intelligibility} {was} considered because the entire data set was not available. Ten of the Harvard sentences and nine of the enhancement algorithms {were used}. The algorithms are referred to in \cite{cooke2013intelligibility} as AdaptDRC, F0-shift, IWFEMD, on/offset, OptimalSII, RESSYSMOD, SBM, SEO, and SSS. In total there are 2 noise types $\times$ 3 SNRs $\times$ (9 algorithms + 1 unprocessed) = 60 conditions.
\subsubsection{KjemsITFS}
The fourth data set consists of speech subjected to ideal time-frequency segregation processing (ITFS). In \cite{kjems2009role} phrases from the Dantale II corpus \cite{wagener2003design} were degraded by four types of noise: SSN, cafeteria noise, noise from a bottling factory, and car noise. For each noise type, the degraded signals were processed by two types of ITFS called an ideal binary mask and a target binary mask. Three SNRs were used ($-60$ dB, and SNRs corresponding to 20\% and 50\% intelligibility) and eight variants of each ITFS algorithm were considered. In total there are 168 conditions. The stimuli were presented to {15} normal-hearing subjects for identification.

\subsection{Performance Measures}
The most important characteristic of an intelligibility metric is that it has a strong monotonic increasing relationship with intelligibility. This paper uses two performance measures to quantify the strength of the relationship: Kendall's tau coefficient \cite{kendall1938new} $\tau$, and Pearson's correlation coefficient $\rho$. To use $\rho$ effectively, the relationship between the metric, $d$, and intelligibility, $w$, must be linear. For this reason, a monotonic function $g(\cdot)$ is applied to $d$ to linearise the relationship:
\begin{equation}
	g(d) = 100(1-e^{-ad})^b,
	\label{mapping}
\end{equation}
where $a,b>0$ are free parameters that are fit to each data set to minimise the mean squared error between $w$ and $g(d)$ over all conditions. These free parameters are affected by the speech corpus, apparatus, and experimental procedures used during the listening test. Pearson's correlation coefficient between $w$ and $g(d)$ is then computed.

\section{Results}
\begin{figure}[!t]
\centering
\includegraphics[scale=1]{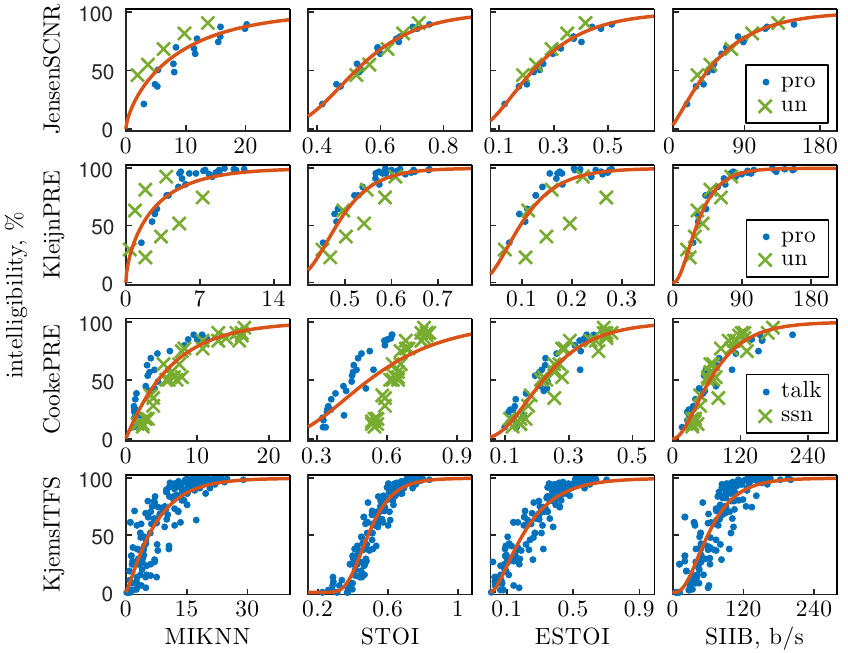}
\caption{Scatter plots of listening test scores (percentage of words correct) against scores computed by intelligibility metrics. For an ideal metric, all points would lie on the fitted curves. Some stimuli involved speech processed by enhancement algorithms (pro) and other stimuli were unprocessed (un). The type of noise in CookePRE is indicated by `talk' and `ssn'.}
\label{fig_scatter}
\end{figure}

The performance of SIIB is compared to three state-of-the-art intelligibility metrics: STOI \cite{taal2011algorithm}, ESTOI \cite{jensen2016algorithm}, and MIKNN \cite{taghia2014objective}. 
Fig. \ref{fig_scatter} shows scatter plots for each data set and each intelligibility metric. The vertical axis shows the intelligibility $w$ and the horizontal axis shows the score computed by an intelligibility metric $d$. Each point represents a different condition in the data set. The function in \eqref{mapping} that is used to linearise the relationship is also shown. Table \ref{table:tau} displays $\tau$ for each data set and metric and, similarly, Table \ref{table:rho} displays $\rho$.

The row of scatter plots corresponding to KleijnPRE shows that all of the reference metrics struggle to predict the effect that optimal energy redistribution has on intelligibility. In contrast SIIB is strongly correlated with intelligibility for this data set ($\tau=0.86$ and $\rho=0.98$).

For CookePRE all of the metrics have reasonable performance except for STOI. This is in agreement with \cite{jensen2016algorithm} which showed that STOI performs poorly for speech degraded by modulated noise sources such as interfering talkers. An assumption sometimes made by the speech processing community is that in order to predict intelligibility for modulated noise sources, statistics have to be averaged over short-time segments to capture the affect of `listening for glimpses of clean speech' \cite{cooke2006glimpsing}. It is then surprising that SIIB performs well on this data set ($\tau=0.76$ and $\rho=0.95$) because SIIB is based on global statistics only.

Compared to the reference metrics SIIB has excellent performance for JensenSCNR, KleijnPRE, and CookePRE, but poorer performance for KjemsITFS ($\tau=0.73$ and $\rho=0.88$). In \cite{taal2011evaluation} seventeen intelligibility metrics were evaluated using KjemsITFS and only five metrics achieved $\rho \ge0.85$. SIIB may not perform as well on KjemsITFS because ITFS processing generates some stimuli with distortions that are not normally encountered in nature. For these stimuli it is plausible that humans are poor decoders. SIIB may correctly estimate the mutual information rate, 
 but humans may be unable to efficiently use all of the information. This hypothesis could be tested by extensively training listeners to decode ITFS processed speech before conducting a listening test. 

Notice that for maximum intelligibility, SIIB estimates an information rate of about $150$ b/s. This is higher than estimates based on linguistic models of speech communication where the information rate is 50-100 b/s \cite{fano1950information,flanagan1972speech,villasenor2012information}. This overestimate is likely the consequence of approximating $\mathbf{X}^K$ as Gaussian. Since $\mathbf{X}^K$ is only approximately Gaussian, the KLT does not remove all statistical dependencies. Accounting for the remaining dependencies would give a lower information rate. 

\begin{table}                                                                   
\centering          
\caption{Performance of intelligibility metrics in terms of Kendall's tau coefficient, $\tau$.}   
\begin{tabular*}{\columnwidth}{lc@{\extracolsep{\fill}}ccc} 
\toprule  \toprule                                                                      
 & MIKNN & STOI & ESTOI & SIIB \\                                          
\midrule                                                                        
JensenSCNR & 0.68 & 0.89 & 0.83 & 0.92 \\                                     
KleijnPRE & 0.71 & 0.70 & 0.58 & 0.86 \\                                      
CookePRE & 0.72 & 0.56 & 0.77 & 0.76 \\                                       
KjemsITFS & 0.71 & 0.82 & 0.81 & 0.73 \\                      
\textbf{Mean}  & \textbf{0.71} &  \textbf{0.75} & \textbf{0.75} &  \textbf{0.82}\\              
\bottomrule  \bottomrule                                                                   
\end{tabular*}                                                                   
\label{table:tau}                                                      
\end{table}        

\begin{table}                                                                           
\centering            
\caption{Performance of intelligibility metrics in terms of Pearson's correlation coefficient, $\rho$.}                                                                  
\begin{tabular*}{\columnwidth}{lc@{\extracolsep{\fill}}ccc}                                                                  
\toprule \toprule                                                                               
 & MIKNN & STOI & ESTOI & SIIB \\                                                  
\midrule                                                                                 
JensenSCNR & 0.86 & 0.99 & 0.98 & 0.99 \\                                     
KleijnPRE & 0.80 & 0.91 & 0.81 & 0.98 \\                                      
CookePRE & 0.90 & 0.69 & 0.95 & 0.95 \\                                       
KjemsITFS & 0.88 & 0.96 & 0.95 & 0.88 \\ 
\textbf{Mean} & \textbf{0.86} & \textbf{ 0.89} & \textbf{0.92} &  \textbf{0.95}\\                                          
\bottomrule      \bottomrule                                                                       
\end{tabular*}                                                                           
\label{table:rho}                                                              
\end{table}

\section{Conclusion}
In this paper we proposed an intrusive instrumental intelligibility metric called SIIB. SIIB is based on the hypothesis that intelligibility is related to the amount of information shared between a clean and degraded speech signal in bits per second. Compared to existing metrics, SIIB is conceptually simple, theoretically motivated, and has high performance. According to Occam's razor, these properties suggest that SIIB might generalise well to new data sets. A MATLAB implementation is available at \url{https://stevenvankuyk.com/matlab_code/}

\vfill

\pagebreak


%


\ifCLASSOPTIONcaptionsoff
  \newpage
\fi



\bibliographystyle{IEEEtran}
\end{document}